\documentclass{article}
\usepackage{amsthm}

\theoremstyle{definition}
\newtheorem{definition}{Definition}
\theoremstyle{plain}
\newtheorem{theorem}{Theorem}
\theoremstyle{remark}
\newtheorem{remark}{Remark}
\theoremstyle{plain}
\newtheorem{lemma}{Lemma}

\begin{document}

\begin{center}

\begin{Large}
\textbf{Geometrical aspects of integrable systems}
\end{Large}
\bigskip

\begin{large}
Emanuele Fiorani
\end{large}

Department of Mathematics and Informatics, University of Camerino \\
62032 Camerino (MC), Italy \\
emanuele.fiorani@unicam.it

\end{center}
\bigskip

\begin{abstract}
We review some basic theorems on integrability of Hamiltonian
systems, namely the Liouville-Arnold theorem on complete
integrability, the Nekhoroshev theorem on partial integrability and
the Mi\-sh\-che\-nko-Fo\-me\-nko theorem on noncommutative
integrability, and for each of them we give a version suitable for
the noncompact case. We give a possible global version of the
previous local results, under certain topological hypotheses on the
base space. It turns out that locally affine structures arise
naturally in this setting.
\bigskip

\end{abstract}

\section{Introduction}

There are some well known theorems treating integrable Hamiltonian
systems on a $2n$-dimensional symplectic manifold $(Z, \Omega )$:
the so called Liouville-Arnold theorem (or
Li\-ou\-vi\-lle-Mi\-ne\-ur-Arno\-ld theorem) for completely
integrable systems (see e.g. \cite{Arnold_B,Libermann_B}), the
Nekhoroshev theorem (or Poin\-car\'{e}-Lya\-pou\-nov-Nek\-horo\-shev
theorem) for partially integrable systems (see
\cite{Nekhoroshev_A}), and the Mi\-sh\-che\-nko-Fo\-me\-nko theorem
for noncommutative (i.e. non-Abelian) systems (see
\cite{Fasso_A,Mishchenko_A}). All of them state the existence of a
neighborhood of a compact invariant submanifold (which is a torus
$\mathbf{T}^n$) which is topologically trivial (i.e. a product), and
the existence of particular adapted coordinates on it, called
action-angle coordinates. These theorems have been extended to the
case of noncompact invariant submanifold (which is a cylinder
$\mathbf{R}^{n-h} \times \mathbf{T}^h$); see
\cite{Fiorani1_A,Fiorani2_A,Fiorani3_A,Giachetta_A}. Sections 2, 3
and 4 intend to compare the compact and noncompact situations.

If invariant submanifolds are compact, topological obstructions to
the global triviality of the whole $Z$ and to the global existence
of action-angle coordinates have been studied (see
\cite{Dazord_A,Duistermaat_A,Nekhoroshev2_A}). Here we aim to extend
this study to the case of noncompact invariant submanifolds; a
possible approach is shown in Section 5 and follows that of
\cite{Fiorani4_A}. The main idea is to provide a sufficient
condition in order that a principal bundle with Abelian structure
group be trivial.

Section 6 is devoted to the interplay between integrability and
locally affine structures. The motivation is to clarify the meaning
of a certain connection naturally arising on the invariant
submanifolds; see for instance \cite{Fasso2_A,Woodhouse_B}. For this
purpose it is needed to study in some details the relation between
flat and parallelizable manifolds and, in this context, the role of
the torsion of a flat connection is considered. Some peculiarity of
spheres and Euclidean spaces are outlined.

Throughout this paper all manifolds are smooth and paracompact and
all maps and structures are smooth, $(Z, \Omega)$ is a
$2n$-dimensional connected symplectic manifold and $(C^{\infty}(Z),
\{ \})$ is the Poisson algebra of (real) functions on $Z$. Given a
function $F: Z \rightarrow \mathbf{R}$, its Hamiltonian vector field
$\Omega^{\sharp}$d$F$ is indicated with $X_F$.

\section{Complete integrability}

Perhaps Theorem \ref{ComplComp} below is the prototype of
integrability theorems; it treats Hamiltonian systems with the
maximum possible number of (independent) functions in involution.
This theorem, together with further discussions, can be found for
instance in the books \cite{Arnold_B,Libermann_B} which are standard
references.

\begin{definition}
A subset $(F_1, \ldots, F_n)$ of $(C^{\infty}(Z), \{ \})$ is called
a completely integrable system on $Z$ if at every point of $Z$
$\mathrm{d}F_1, \ldots, \mathrm{d}F_n$ are linearly independent and
$\{F_i, F_j\} = 0$ $\forall \, i,j = 1, \ldots , n$.
\end{definition}

\begin{theorem} \label{ComplComp}
 Let us assume the following:
 \begin{itemize}
  \item[(i)] $F = (F_1, \ldots, F_n): Z \rightarrow \mathbf{R}^n$ is a
  completely integrable system;

  \item[(ii)] There is a compact component $M$ of a fiber of $F$.
 \end{itemize}

 Then:

 \begin{itemize}
  \item[(I)] There exists a (saturated) neighborhood $U$ of $M$ such that
  $F|_U$ is a trivial principal bundle with structure group the torus
  $\mathbf{T}^n$, and $M$ is one of its fibers;

  \item[(II)] Given standard coordinates $(\varphi^{\lambda})$ on
  $\mathbf{T}^n$, the neighborhood $U$ is provided with Darboux
  coordinates $(I_{\lambda}, \varphi^{\lambda})$ such that
  $$
  \Omega = \mathrm{d}I_{\lambda} \wedge \mathrm{d}\varphi^{\lambda}
  $$
  and $F_{\lambda} = F_{\lambda}(I_{\mu})$ on $U$.
 \end{itemize}
\end{theorem}

\begin{remark}
\begin{itemize}
 \item[(i)] If $F: A \rightarrow B$ is a submersion, the compact components
 of the fibers of $F$ make up an {\em open} subset of $A$; see the
 discussion on page 358 of \cite{Guillemin_B}.

 \item[(ii)] A submersion $F: A \rightarrow B$ ($B$ connected) with
 compact fibers having constant (finite) number of components is a
 bundle; indeed it can be shown that, under these hypotheses, $F$ is
 a proper map. Alternatively, if the fibers of $F$ are diffeomorphic
 to a Euclidean space $\mathbf{R}^n$ then $F$ is a bundle.
 See \cite{Meigniez_A}.

 Thus (i) and (ii) can be used to give a simple proof
 of Theorem \ref{ComplComp}.

 \item[(iii)] In contrast with (ii), a submersion with fibers
 diffeomorphic to a product $\mathbf{R}^n \times K$, where $K$ is a
 compact manifold, need {\em not} be a bundle. See \cite{Giachetta_B},
 Example 1.2.2, in which a submersion onto $\mathbf{R}$ with fibers
 all diffeomorphic to $\mathbf{R} \times \mathbf{S}^1$ is shown
 to have no trivial neighborhood of 0 $\in \mathbf{R}$.
\end{itemize}
\end{remark}

It is possible to remove the compactness hypothesis for $M$ in the
previous theorem in the following way; see
\cite{Fiorani1_A,Fiorani2_A} for a proof and further discussions.

\begin{theorem} \label{ComplNonComp}
 Let us assume the following:
  \begin{itemize}
  \item[(i)] $F = (F_1, \ldots, F_n): Z \rightarrow \mathbf{R}^n$ is a
  completely integrable system;

  \item[(ii)] The fibers of $F$ are connected and mutually diffeomorphic;

  \item[(iii)] Each Hamiltonian vector field $X_{F_{\lambda}}$ is complete.
 \end{itemize}

Then:

 \begin{itemize}
  \item[(I)] $F$ is a bundle with typical fiber the cylinder
  \begin{equation}
   \mathbf{R}^{n-h} \times \mathbf{T}^{h}; \label{CylCompl}
  \end{equation}
  moreover there is a covering of $F(Z)$ with simply connected domains
  $\{V_{\alpha}\}$ such that  over each $V_{\alpha}$ $F$ is
  a principal bundle with structure group (\ref{CylCompl});

  \item[(II)] Given standard coordinates $(y^{\lambda})$ on (\ref{CylCompl}),
  each fiber of $F$ has a (saturated) neighborhood $U$ provided with
  Darboux coordinates $(I_{\lambda}, y^{\lambda})$ such that
  $$
  \Omega = \mathrm{d}I_{\lambda} \wedge \mathrm{d}y^{\lambda}
  $$
  and $F_{\lambda} = F_{\lambda}(I_{\mu})$ on $U$.
 \end{itemize}
\end{theorem}

\begin{remark}
\begin{itemize}
 \item[(i)] If we remove the completeness hypothesis (iii) in Theorem
 \ref{ComplNonComp}, then the fibers of $F$ can be much more complicated than
 $\mathbf{R}^{n-h} \times \mathbf{T}^{h}$: see \cite{Flaschka_A}
 where it is shown that if we take for $F$ a polynomial with two complex
 variables, say $(z_1, z_2) \in \mathbf{C}^2 \simeq \mathbf{R}^4$, then the
 functions Re$(F)$ and Im$(F)$ are in involution with respect to a
 certain symplectic structure on $\mathbf{R}^4$. Thus we can have
 generic Riemann surfaces as fibers of such an $F$.

 \item[(ii)] If one of the fibers of $F$ is compact, then it must be a torus
 and by (i) of Remark 3 there is a saturated neighborhood of it made up
 of such compact fibers; moreover each $X_{F_{\lambda}}$ is complete on them.
 Thus all the hypotheses of Theorem \ref{ComplNonComp} are verified and we
 can see Theorem \ref{ComplComp} as a special case of Theorem
 \ref{ComplNonComp}.
\end{itemize}
\end{remark}

\section{Partial integrability}

Theorem \ref{ParComp} below treats the case of a Hamiltonian system
with a number of (independent) functions in involution lesser than
the maximum possible. It first appeared in the article of N. N.
Nekhoroshev \cite{Nekhoroshev_A} but a more detailed proof of it,
together with discussions and some worked out examples, can be found
in \cite{Bambusi_A,Gaeta_A}. Moreover, Proposition 27.8 of
\cite{Guillemin_B} is a related result: it gives a sufficient
condition for a foliation with a compact leaf to be a trivial bundle
around that leaf.

\begin{definition}
Let $k$ be an integer, $1 \leq k \leq n$.
A subset $(F_1, \ldots, F_k)$ of $(C^{\infty}(Z), \{ \})$ is called
a partially integrable system on $Z$ if at every point of $Z$
$\mathrm{d}F_1, \ldots, \mathrm{d}F_k$ are linearly independent and
$\{F_i, F_j\} = 0$ $\forall \, i,j = 1, \ldots , k$.
\end{definition}

\begin{theorem} \label{ParComp}
 Let us assume the following:
 \begin{itemize}
  \item[(i)] $F = (F_1, \ldots, F_k): Z \rightarrow \mathbf{R}^k$ is a
  partially integrable system;

  \item[(ii)] There is a compact leaf $M$ of the foliation spanned by the
  Hamiltonian vector fields $X_{F_1}, \ldots, X_{F_k}$;

  \item[(iii)] There is a closed curve $\gamma$ on $M$ such that
  D$\Psi_{\gamma}$ does not have 1 as an eigenvalue, where
  $\Psi_{\gamma}$ is the Poincar\'{e} map corresponding to $\gamma$.
 \end{itemize}

Then:

 \begin{itemize}
  \item[(I)] There exists a {\em submanifold} $N$ of $Z$, $N$ of dimension 2$k$
  and $N \supseteq M$, such that $F|_N$ is a trivial principal bundle with
  structure group $\mathbf{T}^n$, and $M$ is one of its fibers;

  \item[(II)] Given standard coordinates $(\varphi^{\lambda})$ on $\mathbf{T}^k$,
  $N$ is provided with Darboux coordinates $(I_{\lambda},
  \varphi^{\lambda})$ such that
  $$
  \Omega = \mathrm{d}I_{\lambda} \wedge \mathrm{d}\varphi^{\lambda}
  $$
  and $F_{\lambda} = F_{\lambda}(I_{\mu})$ on $N$.
 \end{itemize}
\end{theorem}

It is possible to remove the compactness hypothesis for the leaf $M$
in the previous theorem in the following way; see
\cite{Fiorani2_A,Giachetta_A} for a proof and further discussions.

\begin{theorem} \label{ParNonComp}
 Let us assume the following:
 \begin{itemize}
  \item[(i)] $F = (F_1, \ldots, F_k): Z \rightarrow \mathbf{R}^k$ is a
  partially integrable system;

  \item[(ii)] The leaves of the foliation spanned by the Hamiltonian
  vector fields $X_{F_1}, \ldots, X_{F_k}$ are mutually
  diffeomorphic and are the fibers of a submersion
  $\mathcal{F}: Z \rightarrow B$;

  \item[(iii)] Each Hamiltonian vector field $X_{F_{\lambda}}$ is complete.

\end{itemize}

Then:

 \begin{itemize}
  \item[(I)] $\mathcal{F}$ is a bundle with typical fiber the cylinder
  \begin{equation}
   \mathbf{R}^{k-h} \times \mathbf{T}^{h}; \label{CylPar}
  \end{equation}
  moreover there is a covering of $\mathcal{F}(Z)$ with simply connected
  domains $\{V_{\alpha}\}$ such that  over each $V_{\alpha}$ $F$ is
  a principal bundle with structure group (\ref{CylPar});

  \item[(II)] Given standard coordinates $(y^{\lambda})$ on (\ref{CylPar}),
  each fiber of $\mathcal{F}$ has a (saturated) neighborhood $U$
  provided with Darboux coordinates
  $(I_{\lambda}, p_A, q^A, y^{\lambda})$ such that
  $$
   \Omega = \mathrm{d}I_{\lambda} \wedge \mathrm{d}y^{\lambda}
    + \mathrm{d}p_A \wedge \mathrm{d}q^A
  $$
  and $F_{\lambda} = F_{\lambda}(I_{\mu})$ on $U$.
 \end{itemize}
\end{theorem}

\begin{remark}
\begin{itemize}
 \item[(i)] If $k = n$ this theorem reduces to Theorem
 \ref{ComplNonComp} about complete integrability.

 \item[(ii)] The Hamiltonian vector fields $\{X_{I_{\lambda}}
 = \frac{\partial}{\partial y^{\lambda}}\}$ are obviously complete,
 but the Hamiltonian vector fields $\{X_{p_A} = \frac{\partial}{\partial q^A}\}$
 and $\{X_{q^A} = -\frac{\partial}{\partial p_A}\}$ may fail to be
 complete.

 If $\{X_{p_A}\}$ are complete, then the functions $(I_{\lambda}, p_A)$
 verifies the hypotheses of Theorem \ref{ComplNonComp} about
 complete integrability on $U$. Moreover if also $\{X_{q^A}\}$ are complete,
 then the functions $(I_{\lambda}, p_A, q^A)$ verifies the hypotheses
 of Theorem \ref{SupNonComp} below about noncommutative integrability on $U$.

 This is indeed more than it is needed; see part (ii) of Remark
 \ref{RemSup}.
\end{itemize}
\end{remark}

\section{Noncommutative integrability}

In the previous sections we considered subsets of $(C^{\infty}(Z),
\{\})$ of at most $n$ (independent) functions which commute, i.e.
generate a commutative Lie algebra. Now we want to treat the case of
a number of functions greater that $n$; of course in this case they
cannot be mutually commuting (if they are independent).

The following theorem goes in this direction; it first appeared in
the the article of A. S. Mishchenko and A. T. Fomenko
\cite{Mishchenko_A} in the special case in which the functions $F_1,
\ldots, F_k$ generate a certain \emph{finite} dimensional Lie
algebra, in general not commutative (here $n \leq h < 2n$). Theorem
\ref{SupComp} in the present form (which does not need this
restriction) can be found in \cite{Fasso_A}.

\begin{theorem} \label{SupComp}
 Let us assume the following:
 \begin{itemize}
  \item[(i)] $F = (F_1, \ldots, F_k): Z \rightarrow \mathbf{R}^k$ is a
  submersion with compact and connected fibers (here $n \leq k < 2n$);

  \item[(ii)] $\{F_{\lambda}, F_ {\mu}\} = s_{\lambda\mu}(F)$ with the
  $k \! \times \! k$ matrix $(s_{\lambda\mu})$ of constant rank $\, 2(k-n)$.
 \end{itemize}

 Then:

 \begin{itemize}
  \item[(I)] $F$ is a bundle with typical fiber the torus
  $\mathbf{T}^{2n-k}$; moreover there is a covering of $F(Z)$ with
  simply connected domains $\{V_{\alpha}\}$ such that over each
  $V_{\alpha}$ $F$ is a principal bundle with structure group
  $\mathbf{T}^{2n-k}$;

  \item[(II)] Given standard coordinates $(\varphi^i)$ on
  $\mathbf{T}^{2n-k}$, each fiber of $F$ has a (saturated)
  neighborhood $U$ provided with Darboux coordinates
  $(I_i, p_A, q^A, \varphi^i)$ such that
  $$
  \Omega = \mathrm{d}I_i \wedge \mathrm{d}\varphi^i
   + \mathrm{d}p_A \wedge \mathrm{d}q^A
  $$
  and $F_{\lambda} = F_{\lambda}(I_i, p_A, q^A)$ on $U$.

  \noindent Given a Hamiltonian function $\mathcal{H}$ of the
  system, it depends only on coordinates $(I_i)$ on $U$.
 \end{itemize}
\end{theorem}

It is possible to remove the compactness hypothesis for the fibers
of $F$ in the previous theorem in the following way; the proof given
here is an improvement of the proof that can be found in
\cite{Fiorani3_A} together with further discussions and a worked out
example.

\begin{theorem} \label{SupNonComp}
 Let us assume the following:
 \begin{itemize}
  \item[(i)] $F = (F_1, \ldots, F_k): Z \rightarrow \mathbf{R}^k$ is a
  submersion with connected and mutually diffeomorphic fibers
  (here $n \leq k < 2n$);

  \item[(ii)] $\{F_{\lambda}, F_ {\mu}\} = s_{\lambda\mu}(F)$ with the
  $k \! \times \! k$ matrix $(s_{\lambda\mu})$ of constant rank $\, 2(k-n)$.

  \item[(iii)] Each Hamiltonian vector field $X_{F_{\lambda}}$ is complete.
 \end{itemize}

 Then:

 \begin{itemize}
  \item[(I)] $F$ is a bundle with typical fiber the cylinder
  \begin{equation}
   \mathbf{R}^{2n-k-h} \times \mathbf{T}^{h}; \label{CylNonComm}
  \end{equation}
  moreover there is a covering of $F(Z)$ with simply connected
  domains $\{V_{\alpha}\}$ such that over each $V_{\alpha}$ $F$ is a
  principal bundle with structure group (\ref{CylNonComm}).

  \item[(II)] Given standard coordinates $(y^i)$ on
  (\ref{CylNonComm}), each fiber of $F$ has
  a (saturated) neighborhood $U$ provided with Darboux coordinates
  $(I_i, p_A, q^A, y^i)$ such that
  $$
  \Omega = \mathrm{d}I_i \wedge \mathrm{d} y^i
   + \mathrm{d}p_A \wedge \mathrm{d}q^A
  $$
  and $F_{\lambda} = F_{\lambda}(I_i, p_A, q^A)$ on $U$.

  \noindent Given a Hamiltonian function $\mathcal{H}$ of the
  system, it depends only on coordinates $(I_i)$ on $U$.
 \end{itemize}
\end{theorem}

\begin{proof}
$F$ is a submersion, so $F(Z)$ is an open subset of $\mathbf{R}^k$
and we can choose standard coordinates $(x_{\lambda})$ on it.

Since $\forall \: \lambda, \mu = 1, \ldots k$ we have
$\{F_{\lambda}, F_{\mu}\} = s_{\lambda\mu}(F)$, it is possible to
provide $F(Z)$ with a Poisson structure $\{\}_F$ whose Poisson
tensor is just
\begin{equation} \label{Poisson}
s_{\lambda\mu}(x) \frac{\partial}{\partial x_{\lambda}} \wedge
\frac{\partial}{\partial x_{\mu}}
\end{equation}
and $F$ is a Poisson morphism.

Since $\forall \, x \! \in \! F(Z)$ we have rank$(s_{\lambda\mu}(x))
= 2(k-n)$, around each $x \in F(Z)$ there are $2n-k$ independent
Casimir functions $\{C_i\}$ of $\{\}_F$ such that the Hamiltonian
vector fields $\{X_{F^*C_i}\}$ span the fibers of $F$. See
\cite{Libermann_B} for these facts, where $\{\}_F$ is called
\emph{coinduced Poisson structure}.

Now if we consider the Hamiltonian vector field
$\{X_{F_{\lambda}}\}$ instead of the functions $\{F_{\lambda}\}$,
from (ii) it follows that
 $$
[X_{F_{\lambda}}, X_{F_{\mu}}] = - \frac{\partial
s_{\lambda\mu}}{\partial x_{\alpha}} X_{F_{\alpha}}
 $$
i.e. on each fiber of $F$ they generate a \emph{finite} dimensional
Lie algebra; then each linear combination of $\{X_{F_{\lambda}}\}$
is complete; see \cite{Palais_A}. In particular, each Hamiltonian
vector field of the functions $\{F^*C_i\}$
\begin{equation} \label{LinCombFields}
X_{F^*C_i} = \frac{\partial C_i}{\partial x_{\alpha}} X_{F_{\alpha}}
\end{equation}
is complete.

Then it is possible to apply the previous Theorem \ref{ParNonComp}
to the partial integrable system on $Z$ given by $\{F^*C_i\}$.
\end{proof}

\begin{remark} \label{RemSup}
\begin{itemize}
 \item[(i)] If $k = n$ this theorem reduces to Theorem
 \ref{ComplNonComp} about complete integrability.

 \item[(ii)] The Hamiltonian vector fields $\{X_{I_i}
 = \frac{\partial}{\partial y^i}\}$ are obviously complete,
 thus the functions $(I_i)$ verifies the hypotheses of Theorem
 \ref{ParNonComp} about partial integrability on $U$ if we take
 as $\mathcal{F}$ the projection $(I_i, p_A, q^A, y^i)
 \mapsto (I_i, p_A, q^A)$.

 The Hamiltonian vector fields $\{X_{p_A} = \frac{\partial}{\partial q^A}\}$
 and $\{X_{q^A} = -\frac{\partial}{\partial p_A}\}$ may fail to be
 complete. If $\{X_{p_A}\}$ are complete, then the functions
 $(I_i, p_A)$ verifies the hypotheses of Theorem \ref{ComplNonComp}
 about complete integrability on $U$.

 In any case, the completeness hypothesis (iii) here is used only to
 guarantee that each field $X_{F^*C_i}$ in (\ref{LinCombFields}) is
 complete.
\end{itemize}
\end{remark}

\section{Integrability and global action-angle coordinates}

So far all the theorems stated the existence of action-angle
coordinates on a (saturated) neighborhood of any invariant manifold.
It is natural now to investigate the global picture: for the compact
case standard references are the articles of P. Dazord and T.
Delzant \cite{Dazord_A}, J. J. Duistermaat \cite{Duistermaat_A} and
N. N. Nekhoroshev \cite{Nekhoroshev2_A}. In what follows there is a
possible way to treat the case of noncompact invariant manifolds;
Theorem \ref{GlobPar}, \ref{GlobSup} and \ref{GlobCompl} can be
found in \cite{Fiorani4_A}; let us note that Lemma \ref{LemmaGlob}
and Theorem \ref{PrincGlob} here establish a basic sufficient
condition for a principal bundle with Abelian structure group to be
trivial, and permit to simplify the proofs of the theorems contained
in \cite{Fiorani4_A}.

\begin{lemma} \label{LemmaGlob}
Let $P \rightarrow B$ be a principal bundle and let us assume that
its structure group is a product $G_1 \times G_2$.

Then it is canonically isomorphic to the product principal bundle
$P_1 \times P_2 \rightarrow B$ where $P_1 \rightarrow B$ is the
quotient principal bundle $P/G_2 \rightarrow B$ and $P_2 \rightarrow
B$ is $P/G_1 \rightarrow B$.
\end{lemma}

\begin{proof}
Using the fact that we have canonical projections of $G_1 \times
G_2$ onto its factors $G_1$ and $G_2$, the proof is a
straightforward check.
\end{proof}

\begin{theorem} \label{PrincGlob}
Let $P \rightarrow B$ be a principal bundle and let us assume the
following:
\begin{itemize}
 \item[(i)] The structure group $G$ is Abelian;

 \item[(ii)] The base $B$ verifies $H^2(B, \textbf{Z})=0$.
\end{itemize}
Then the principal bundle $P \rightarrow B$ is trivial.
\end{theorem}

\begin{proof}
Since the structure group $G$ is Abelian, it is isomorphic to
$\textbf{R}^n$ or to the cylinder $\textbf{R}^{n-h} \times \textbf{T}^h$,
where $n =\,$dim$G$ and $h \geq 1$. In the first case, the result
follows immediately from the existence of a global section;
see \cite{Kobayashi_B}.

In the second case, it is possible to reduce the group
$\textbf{R}^{n-h} \times \textbf{T}^h$ to its maximal compact
subgroup $\textbf{T}^h$; then, by Lemma \ref{LemmaGlob}, it is
isomorphic to the product of $h$ copies of principal $U(1)$-bundles
over the same base $B$. Since $H^2(B, \textbf{Z})=0$, each of these
bundles has trivial first Chern class and consequently it is
trivial; see \cite{Griffiths_B}. Thus the original principal bundle
is trivial.
\end{proof}

The following theorems are global versions of Theorem
\ref{ParNonComp} and \ref{SupNonComp} about partial and
noncommutative integrability.

\begin{theorem} \label{GlobPar}
 Let us assume the following:
 \begin{itemize}
  \item[(i)] $F = (F_1, \ldots, F_k): Z \rightarrow \mathbf{R}^k$ is a
  partially integrable system;

  \item[(ii)] The leaves of the foliation spanned by the Hamiltonian
  vector fields $X_{F_1}, \ldots, X_{F_k}$ are mutually
  diffeomorphic and are the fibers of a submersion
  $\mathcal{F}: Z \rightarrow B$;

  \item[(iii)] Each Hamiltonian vector field $X_{F_{\lambda}}$ is
  complete;

  \item[(iv)] The base $B$ is simply connected and $H^2(B,
  \mathbf{Z}) = 0$.
\end{itemize}

Then:

 \begin{itemize}
  \item[(I)] $\mathcal{F}$ is a trivial principal bundle with structure group
  the cylinder $\mathbf{R}^{k-h} \times \mathbf{T}^{h}$;

  \item[(II)] Given standard coordinates $(y^{\lambda})$ on
  $\mathbf{R}^{k-h} \times \mathbf{T}^{h}$, $Z$ is provided with
  coordinates $(I_{\lambda}, x^A, y^{\lambda})$ such that
  $$
   \Omega = \mathrm{d}I_{\lambda} \wedge \mathrm{d}y^{\lambda}
    + \Omega^{\lambda}_A \mathrm{d}I_{\lambda} \wedge \mathrm{d}x^A
    + \Omega_{AB} \mathrm{d}x^A \wedge \mathrm{d}x^B
  $$
  and $F_{\lambda} = F_{\lambda}(I_{\mu})$ on $Z$.
 \end{itemize}
\end{theorem}

\begin{theorem} \label{GlobSup}
 Let us assume the following:
 \begin{itemize}
  \item[(i)] $F = (F_1, \ldots, F_k): Z \rightarrow \mathbf{R}^k$ is a
  submersion with connected and mutually diffeomorphic fibers
  (here $n \leq k < 2n$);

  \item[(ii)] $\{F_{\lambda}, F_ {\mu}\} = s_{\lambda\mu}(F)$ with the
  $k \! \times \! k$ matrix $(s_{\lambda\mu})$ of constant rank $\, 2(k-n)$.

  \item[(iii)] Each Hamiltonian vector field $X_{F_{\lambda}}$ is
  complete;

  \item[(iv)] There is an open subset $V$ of $F(Z)$ admitting $2n-k$
  independent Casimir functions of the coinduced Poisson structure on $F(Z)$
  as in the proof of Theorem \ref{SupNonComp} which is simply connected
  and such that $H^2(V, \mathbf{Z}) = 0$.
 \end{itemize}

Then:

 \begin{itemize}
  \item[(I)] The restriction $F|V: F^{-1}(V) \rightarrow V$ is a trivial
  principal bundle with structure group the cylinder
  $\mathbf{R}^{2n-k-h} \times \mathbf{T}^{h}$;

  \item[(II)] Given standard coordinates $(y^i)$ on
  $\mathbf{R}^{2n-k-h} \times \mathbf{T}^{h}$, $F^{-1}(V)$ is provided with
  coordinates $(I_{\lambda}, x^A, y^{\lambda})$ such that
  $$
   \Omega = \mathrm{d}I_{\lambda} \wedge \mathrm{d}y^{\lambda}
    + \Omega_{AB} \mathrm{d}x^A \wedge \mathrm{d}x^B
  $$
  and $F_{\lambda} = F_{\lambda}(I_{\mu}, x^A)$ on $F^{-1}(V)$.

  \noindent Given a Hamiltonian function $\mathcal{H}$ of the
  system, it depends only on coordinates $(I_{\lambda})$ on $F^{-1}(V)$.
 \end{itemize}
\end{theorem}

\begin{remark} The proof of Theorem \ref{SupNonComp} shows that
each $x \in F(Z)$ has a neighborhood $V$ admitting $2n-k$
independent Casimir functions; thus the result holds over any such
$V$ which is simply connected and such that $H^2(V, \mathbf{Z}) =
0$.
\end{remark}

The following theorem is a global version of Theorem
\ref{ComplNonComp}; note that in this case the set of hypotheses is
simpler than in the previous cases in that it requires only
topological properties on the image $F(Z)$ of the original
submersion $F$.

\begin{theorem} \label{GlobCompl}
 Let us assume the following:
 \begin{itemize}
  \item[(i)] $F = (F_1, \ldots, F_n): Z \rightarrow \mathbf{R}^n$ is a
  completely integrable system;

  \item[(ii)] The fibers of $F$ are connected and mutually diffeomorphic;

  \item[(iii)] Each Hamiltonian vector field $X_{F_{\lambda}}$ is
  complete;

  \item[(iv)] $F(Z)$ is simply connected and $H^2(F(Z),
  \mathbf{Z}) = 0$.
 \end{itemize}

 Then:

 \begin{itemize}

  \item[(I)] $F$ is a {\rm trivial} principal bundle with structure group
  the cylinder $\mathbf{R}^{n-h} \times \mathbf{T}^{h}$;

  \item[(II)] Given standard coordinates $(y^{\lambda})$ on
  $\mathbf{R}^{n-h} \times \mathbf{T}^{h}$, $Z$ is provided
  with Darboux coordinates $(I_{\lambda}, y^{\lambda})$ such that
  $$
  \Omega = \mathrm{d}I_{\lambda} \wedge \mathrm{d}y^{\lambda}
  $$
  and $F_{\lambda} = F_{\lambda}(I_{\mu})$ on $Z$.
 \end{itemize}
\end{theorem}

\begin{remark} Since in hypothesis (iv) we assume $F(Z)$
to be simply connected, $H^2(F(Z), \mathbf{Z}) = 0$ is equivalent to
$\pi_2(F(Z)) = 0$, i.e. (iv) can be restated by requiring $F(Z)$ to
be 2-connected. The same holds for the base manifold $B$ in Theorem
\ref{GlobPar} and for the neighborhood $V$ in Theorem \ref{GlobSup}.
\end{remark}

\section{Integrability and locally affine structures}

In what follows $M$ is a connected and $m$-dimensional manifold; we
recall the following two standard definitions.

\begin{definition}
Let $\nabla$ be a linear connection on $M$ and let $R_{\nabla}$,
$T_{\nabla}$ be its curvature and torsion tensors respectively;
\begin{itemize}
 \item $\nabla$ is flat when $R_{\nabla} = 0$;

 \item $\nabla$ is a locally affine structure on $M$ when $R_{\nabla} =
 0$ and $T_{\nabla} = 0$.
\end{itemize}
\end{definition}

\begin{definition}
Let $\{X_1, \dots, X_m\}$ be a parallelization of $M$ and let $X$,
$Y$ be vector fields on $M$, $Y = b^jX_j$ with $b^j$ global smooth
functions on $M$. The expression
\begin{equation}
\nabla_XY = X(b^j)X_j \label{ConnPar}
\end{equation}
is easily seen to define a linear connection on $M$, called the
connection of the parallelization.
\end{definition}

The basic properties of the connection (\ref{ConnPar}) are
summarized in the following theorem.

\begin{theorem} \label{ProprConnPar}
Let $\{X_1, \dots, X_m\}$ be a parallelization of $M$ and let
$\nabla$ be as in (\ref{ConnPar}). Then:
\begin{itemize}
 \item[(I)] $\nabla$ is a flat linear connection;

 \item[(II)] $T_{\nabla}(X_i, X_j) = -[X_i, X_j]$;

 \item[(III)] Each $X_i$ is a geodesic vector fields for $\nabla$,
 i.e. each integral curve of $X_i$ is a geodesic of $\nabla$;

 \item[(IV)] Given another parallelization $\{\bar{X}_1, \ldots, \bar{X}_m\}$ of
 $M$, it defines the same $\nabla$ if and only if  $\bar{X}_j = G^i_jX_i$
 $\forall \ j = 1, \ldots, m$, with $(G^i_j)$ some constant
 invertible matrix.
\end{itemize}
\end{theorem}

\begin{proof}
Statement (I) can be found in \cite{Greub_B}, but they all follow
from direct computations. In particular, to prove (IV) it is
sufficient to consider two parallelizations $\{X_i\}$ and
$\{\bar{X}_j\}$ of $M$ and observe that there must be a relation
$\bar{X}_j = G^i_j(x)X_i$ between the two, with $(G^i_j(x))$ an
invertible matrix of smooth functions on $M$. Then the condition
that their associated connections $\nabla$ and $\bar{\nabla}$ must
be equal leads to the equivalent condition $X_k(G^i_j(x)) = 0$
$\forall i,j,k = 1, \ldots, m$ and $\forall x \in M$; the facts that
$\{X_i\}$ is a basis at every point and that $M$ is connected yields
that the matrix $(G^i_j)$ must be constant.
\end{proof}

The following theorem states that when $M$ is simply connected also
the converse of Theorem \ref{ProprConnPar} holds.

\begin{theorem} \label{ConvProprConnPar}
Let us assume the following:
\begin{itemize}
 \item[(i)] There is a flat linear connection $\nabla$ on $M$;

 \item[(ii)] $M$ is simply connected.
\end{itemize}
Then there is a parallelization $\{X_1, \ldots, X_m\}$ of $M$ such
that its connection is $\nabla$.
\end{theorem}

\begin{proof}
The proof can be found in \cite{Greub_B}, but also a direct proof
can be given. Let us fix a point $x_0 \in M$ and a basis $\{X_1,
\ldots, X_m\}$ of $T_{x_0}M$. Since $\nabla$ is flat and $M$ simply
connected, it is possible, through parallel transport, to move this
basis to any point $x \in M$ getting a basis of $T_xM$ and the
result does not depend on the path followed. Thus we get a
parallelization of $M$ still indicated with $\{X_1, \ldots, X_m\}$;
it can be seen that it is smooth and $\nabla_{X_i} X_j = 0$ $\forall
\ i, j = 1, \ldots, m$. It follows that, given any vector field $Y =
b^jX_j$ on $M$ with $b^j$ global smooth functions on $M$, we get for
the original connection $\nabla$ the formula $\nabla_{X_i}Y =
X_i(b^j)X_j$, which coincides with the formula (\ref{ConnPar}) for
the connection of the parallelization $\{X_1, \ldots, X_m\}$.
\end{proof}

\begin{remark} Let us consider the following relation on the set of
all the parallelizations of a manifold $M$: two parallelizations
$\{X_1\, \ldots, X_m\}$ and $\{\bar{X}_1, \ldots, \bar{X}_m\}$ are
equivalent if and only if there is a constant invertible matrix
$(G^i_j)$ such that
\begin{equation} \label{Parallelization}
\bar{X}_j = G^i_jX_i \quad \forall j = 1, \ldots, m.
\end{equation}
It is clear that (\ref{Parallelization}) is an equivalence relation.
From Theorem \ref{ProprConnPar} and \ref{ConvProprConnPar} it
follows that on a \emph{simply connected} manifold $M$, flat linear
connections $\nabla$ are in 1-to-1 correspondence with equivalence
classes of (\ref{Parallelization}).
\end{remark}

We can improve Theorem \ref{ProprConnPar} and \ref{ConvProprConnPar}
by considering the torsion of $\nabla$.

\begin{theorem} \label{ImprProprConnPar}
Let $\{X_1, \ldots, X_m\}$ be a parallelization of $M$ with
commuting vector fields and let $\nabla$ be as in (\ref{ConnPar});
then $\nabla$ is a locally affine structure on $M$.

Conversely, let $\nabla$ be a locally affine structure on $M$ simply
connected; then there is a parallelization $\{X_1, \ldots, X_m\}$ of
$M$ with commuting vector fields such that its connection is
$\nabla$.
\end{theorem}

\begin{proof}
It all follows from Theorem \ref{ProprConnPar} and
\ref{ConvProprConnPar}, remembering the formula $T_{\nabla}(X_i,
X_j) = -[X_i, X_j]$.
\end{proof}

\begin{remark} \label{RemCompl}
\begin{itemize}

 \item[(i)] Let us consider $\{X_1, \ldots, X_m\}$ a parallelization of $M$
 with commuting and {\em complete} vector fields: in this case,  besides
 the locally affine structure (\ref{ConnPar}), we can obtain information
 about $M$. Indeed, the composition $\Phi$ of the flows of the vector fields
 $\{X_1, \ldots, X_m\}$ define a transitive action of $\mathbf{R}^m$
 on $M$. Thus $M$ must be diffeomorphic to $\mathbf{R}^m$ modulo the
 discrete isotropy group of one of its point, which yields
 $\mathbf{R}^{m-h} \times \mathbf{T}^h$ for some $h \in \{0, \ldots,
 m\}$. See for instance \cite{Arnold_B} for this standard argument.
 This situation occurred in in particular in all the theorems of the
 preceding sections.

 \item[(ii)] Of course, also nonorientable manifolds can admit a
 locally affine structure: examples are the M\"{o}bius strip
 and the Klein bottle; they obviously are not simply connected.

\end{itemize}
\end{remark}

It is known that among all the spheres $\mathbf{S}^m$, the
parallelizable ones are $\mathbf{S}^1$, $\mathbf{S}^3$ and
$\mathbf{S}^7$. Putting together the facts of this section, we
obtain the following additional properties; we consider only the
case $\mathbf{S}^m$ with $m \geq 2$, the case $\mathbf{S}^1$ being
trivial.

\begin{theorem} \label{Spheres}
Let us consider the spheres $\mathbf{S}^m$, $m \geq 2$.
\begin{itemize}
 \item[(I)] The only spheres admitting a flat linear connection are
 $\mathbf{S}^3$ and $\mathbf{S}^7$;

 \item[(II)] No sphere $\mathbf{S}^m$ admits a locally
 affine structure.
\end{itemize}
\end{theorem}

\begin{proof} $\mathbf{S}^3$ and $\mathbf{S}^7$ are parallelizable,
thus they admit a flat linear connection. Moreover, each
$\mathbf{S}^m$, $m \geq 2$, is simply connected but for $m \neq 3,
7$ \ they are not parallelizable; thus for $m \neq 3, 7$ \ they
cannot admit a flat linear connection. This proves (I).

According to (I), there remain only $\mathbf{S}^3$ and
$\mathbf{S}^7$ to be checked. Since they are simply connected, if
they admitted a locally affine structure they would be
parallelizable with commuting vector fields. Since they are compact,
these vector fields would be complete and by (i) of Remark
\ref{RemCompl} they would be diffeomorphic to $\mathbf{T}^3$ and
$\mathbf{T}^7$ respectively, which is impossible because they are
not simply connected. This proves (II).
\end{proof}

We can also obtain the following version of the Cartan-Hadamard
theorem:

\begin{theorem} \label{CartanAdamard}
Let us assume the following:
\begin{itemize}
 \item[(i)] There is a locally affine structure $\nabla$ on $M$;

 \item[(ii)] $\nabla$ is geodesically complete, i.e. all of
 its geodesics can be defined on $\mathbf{R}$;

 \item[(iii)] $M$ is simply connected.
\end{itemize}
Then there is a diffeomorphism $\Psi: M \rightarrow \mathbf{R}^m$
such that $\nabla$ corresponds under $\Psi$ to the standard locally
affine structure of \ $\mathbf{R}^m$.
\end{theorem}

\begin{proof}
By Theorem \ref{ImprProprConnPar} $M$ is parallelizable with
commuting vector fields, which are complete by (III) of Theorem
\ref{ProprConnPar}; then, after we fix a starting point $\alpha \in
M$, the map $\Phi: \mathbf{R}^m \rightarrow M$ of (i) of Remark
\ref{RemCompl} must define a diffeomorphism because $M$ is simply
connected. Moreover a computation shows that if $\Phi: t \mapsto p$
then
$$
T_t\Phi: \mbox{e}_i \longmapsto X_i(p) \quad \forall \ i=1, \dots, m
$$
where $(\mbox{e}_1, \dots, \mbox{e}_m)$ is the standard basis of
$\mathbf{R}^m$. Thus it is enough to take as $\Psi$ the function
$\Phi^{-1}$.
\end{proof}

\begin{remark}
\begin{itemize}
 \item[(i)] More generally, we can state the following: no compact
 and simply connected manifold admits a locally affine structure;
 the proof is exactly the same as that of part (II) of Theorem
 \ref{Spheres}. In particular, there we implicitly considered the
 standard differentiable structure of $\mathbf{S}^7$, but the result
 of part (II) holds for all its exotic differentiable structures.

 \item[(ii)] From Theorem \ref{CartanAdamard} it follows that
 on $\mathbf{R}^m$,  after a diffeomorphism, all the geodesically complete
 locally affine structures can be brought to the standard one.
 In particular, since there we implicitly considered the standard
 differentiable structure for $\mathbf{R}^4$, no exotic
 differentiable structure on $\mathbf{R}^4$ admits a
 geodesically complete locally affine structure.
\end{itemize}
\end{remark}

It is well known that locally affine structures arise naturally with
certain foliations of a symplectic manifold. We recall only the
basic definition and construction; for further discussions we refer
to \cite{Dazord_A,Fasso_A,Fasso2_A,Woodhouse_B} and references
therein for the following terminology and contruction.

\begin{definition}
Let $\mathcal{F}$ be a foliation of the symplectic manifold $(Z, \Omega)$.

The polar foliation of $\mathcal{F}$ is, if it exists, the foliation
$\mathcal{F}^{\perp}$ of $Z$ with the property that the tangent spaces
of its leaves are the symplectic complements of the tangent spaces
of the leaves of $\mathcal{F}$.

A foliation $\mathcal{F}$ which admits a polar foliation
$\mathcal{F}^{\perp}$ is called symplectically complete.
\end{definition}

If we have a symplectically complete foliation $\mathcal{F}$ of $Z$
with isotropic leaves, then these leaves admit a locally affine
structure, which is given by
\begin{equation} \label{OmegaComm}
\nabla^{\Omega}_XY = \Omega^{\sharp} L_X \Omega^{\flat}(Y)
\end{equation}
where $X$, $Y$ are vector fields tangent to the leaves of $\mathcal{F}$.

\begin{remark} \label{RemFibers}
It is readily observed that the fibers of the following submersions
\begin{itemize}
 \item[(i)] $F$ of Theorem \ref{ComplNonComp}

 \item[(ii)] $\mathcal{F}$ of Theorem \ref{ParNonComp}

 \item[(iii)] $F$ of Theorem \ref{SupNonComp}
\end{itemize}
define foliations of $Z$ which are isotropic and symplectically complete,
so they admit the locally affine structure $\nabla^{\Omega}$
defined in (\ref{OmegaComm})
\end{remark}

It is natural to investigate the nature of the connection
$\nabla^{\Omega}$ for the cases listed in Remark \ref{RemFibers};
the following theorem states that it is nothing else than the
connection of appropriate parallelizations.

\begin{theorem}
Let us consider the submersions (i), (ii) and (iii) of Remark
\ref{RemFibers} and the connection $\nabla^{\Omega}$ defined on
their fibers by (\ref{OmegaComm}); $\nabla^{\Omega}$ coincides with
the connection $\nabla$ (\ref{ConnPar}) of the following
parallelizations:
\begin{itemize}
 \item For (i) the parallelization given by the
 Hamiltonian vector fields $\{X_{F_1}, \ldots, X_{F_n}\}$;

 \item For (ii) the parallelization given by the
 Hamiltonian vector fields $\{X_{F_1}, \ldots, X_{F_k}\}$;

 \item For (iii) the parallelization given by the
 Hamiltonian vector fields $\{X_{F^*C_1}, \ldots, X_{F^*C_{2n-k}}\}$
 found at the end of the proof of Theorem \ref{SupNonComp}.
\end{itemize}
\end{theorem}

\begin{proof}
To simplify the notations, let us call generically $X_i$ the
Hamiltonian vector field of the function $F_i$; then this proof will
adapt to all the three previous statements. Given a vertical vector
$Y = b^jX_j$, we have
\begin{eqnarray*}
 \nabla^{\Omega}_{X_k}Y & = & \Omega^{\sharp}L_{X_k}\Omega^{\flat}(b^jX_j) =
 \Omega^{\sharp}L_{X_k}(b^j\mbox{d}F_j) = \\
 & = & \Omega^{\sharp}[i_{X_k}\mbox{d}(b^j\mbox{d}F_j)] =
 \Omega^{\sharp}[i_{X_k}(\mbox{d}b^j \wedge \mbox{d}F_j)] = \\
 & = & \Omega^{\sharp}[X_k(b^j)\mbox{d}F_j] = X_k(b^j)X_j = \\
 & = & \nabla_{X_k}Y
\end{eqnarray*}
where we have used that in our case $\Omega(X_i, X_j) = \{F_i, F_j\}
= 0$.
\end{proof}

\end{document}